\documentclass[aps,prl,twocolumn,showpacs]{revtex4}
\usepackage{bm}
\usepackage{graphicx}
\usepackage{amsmath}
\usepackage{eufrak}
\usepackage{color}
\newcommand{\nix}[1]{}
\begin{document}

\title{Helicity sensitive terahertz radiation detection by \\
dual-grating-gate  high electron mobility transistors
}
\author{P. Faltermeier,$^1$ P. Olbrich,$^1$  W. Probst,$^1$ L. Schell,$^1$ 
T. Watanabe$^2$, S. A. Boubanga-Tombet$^2$, T. Otsuji$^2$, 
and S. D.\,Ganichev$^{1}$}
\affiliation{$^1$  Terahertz Center, University of Regensburg, 93040
Regensburg, Germany}
\affiliation{$^2$ Research Institute of Electrical Communication, Tohoku University, 980-8577 Sendai, Japan}

\begin{abstract}
We report on  the observation of a radiation helicity sensitive photocurrent
excited by terahertz (THz) radiation in dual-grating-gate (DGG) InAlAs/InGaAs/InAlAs/InP 
high electron mobility transistors (HEMT). For a circular polarization the current measured 
between source and drain contacts changes its sign with the inversion of the radiation helicity. 
For elliptically polarized radiation the total current is described by superposition 
of the Stokes parameters with different weights. 
Moreover, by variation of gate voltages applied to individual gratings the photocurrent can 
be defined either by the Stokes parameter defining the radiation helicity or those for linear polarization. 
We show that artificial non-centrosymmetric  microperiodic structures with 
a two-dimensional electron system excited by THz radiation exhibit a
$dc$ photocurrent caused by the combined action of a spatially 
periodic in-plane potential and  spatially modulated light. 
The results provide a proof of principle for the application of 
DGG HEMT for all-electric detection of the radiation's polarization state.

\end{abstract}

\maketitle

\section{I. Introduction}

Field-effect-transistors (FETs) have emerged as promising devices 
for sensitive and fast room temperature detection of terahertz  
(THz) radiation~\cite{knaprev2013,knaptredicuccirev2013}.
They are considered as a good candidate for 
real-time THz imaging and spectroscopic analysis~\cite{Boppel2012,Muravev2012} 
as well as future THz wireless communications~\cite{Tonouchi2007}. 
Devices employing plasmonic effects in FETs have already been  
applied  for room 
temperature detection of radiation with frequencies from tens of GHz up to several 
THz and enable the combination of individual detectors in a matrix. 
They are characterized by high responsivity 
(up to a few kV/W), low noise equivalent power 
(down to 10 pW/$\sqrt{\rm Hz}$),  fast response time (tens of picoseconds) 
 and large dynamic range (linear power response  up to 10 kW/cm$^2$), 
see e.g. Ref.\cite{knaptredicuccirev2013,9,14Roj,11,Watanabe,But2014,Knap2014}. 
The  operation principle of FET  THz detectors used so far   is based on the nonlinear 
properties of the two-dimensional (2D) plasma in the transistor channel. 
The standard Dyakonov-Shur model~\cite{Dyakonov} assumes 
that radiation is coupled to the transistor by an effective antenna,
which generates an $ac$ voltage predominantly on one 
side of the transistor. Both resonant~\cite{98} and non-resonant~\cite{99} 
regimes of THz detection have been studied. 
While research aimed to development of THz FET detectors is
focused on single gate structures  recently
several groups have shown that higher sensitivities are expected for structures with periodic symmetric and asymmetric
 metal stripes or gates~\cite{Watanabe,19w,116,20w,16w,18w,31,115,35,31Roj}. 
In particular, dual-grating-gate FET are considered
as a good candidate for sensitive THz detection. 
The first data obtained on dual-gated-structures demonstrated a 
substantial enhancement of the photoelectric response and an ability to control 
detector parameters by variation of individual gate 
bias voltage~\cite{Watanabe}.
At the same time, THz electric field applied to FETs with  asymmetric periodic dual gate structure is expected to give rise to electronic ratchet effects~\cite{28,30,29,31Roj,33Roj} (for review see~\cite{30})  
and plasmonic  ratchet effects~\cite{36}. 
Besides improving the figure of merits of FET detectors,
ratchet effects may also result in  new functionalities. 
In particularly, they may induce photocurrents 
driven solely by the radiation helicity.

Here, we report  on the observation of a radiation helicity sensitive photocurrent
excited by THz radiation in dual-grating-gate InAlAs/InGaAs/InAlAs/InP 
high electron mobility transistors (HEMT). 
We show that artificial non-centrosymmetric 
microperiodic structures with a two-dimensional electron system 
excited by THz radiation exhibit a $dc$ photocurrent caused by the 
lateral asymmetry of the applied static  potential and terahertz electric field. 
We demonstrate that depending on gate voltages applied 
to the individual gratings of the dual-grating-gate the response can be
proportional to either the Stokes parameters~\cite{Saleh} defining the radiation helicity or those for linear polarization.
As an important result, for a wide range of gate voltages we observed a
photocurrent $j_{\rm C}$ being proportional to the radiation helicity $P_{\rm circ} = (I_{\sigma^+} - I_{\sigma^-})/(I_{\sigma^+} + I_{\sigma^-})$, where $I_{\sigma^+}$ and $I_{\sigma^-}$ are intensities of right- and left-handed circularly polarized light.
For the  circular photocurrent $j_{\rm C}$
measured between source and drain contacts changes 
its sign with the inversion of the radiation helicity. 
This observation is of particular importance for 
a basic understanding of plasmon-photogalvanic and 
quantum ratchet effects. It also 
has a large potential 
for the development of 
an all-electric detector of the radiation's polarization state, which was so far realized applying less sensitive photogalvanic effects only~\cite{122,123,124}.
The observed phenomena is discussed in the framework of  
electronic ratchet~\cite{35,30,29,33Roj,31Roj} and plasmonic ratchet 
effects excited in a 2D electron system with a
spatially periodic \textit{dc} in-plane  potential~\cite{Watanabe,35,36}.

\section{II. Experimental Technique}

 %
 %
 %
 %
 %

\begin{figure}
\includegraphics[width=0.80\linewidth]{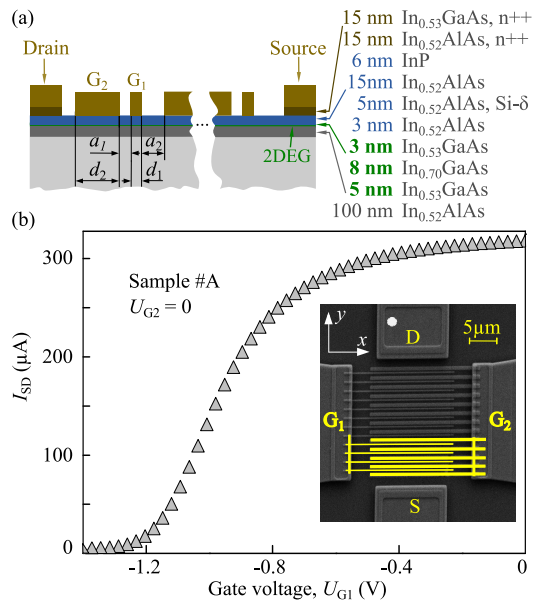}
\caption{(a) 
Sketch of the dual-grating-gate HEMT. 
Cross-section of the structure shows the 
layer sequence and indicates the width of the fingers
($d\rm_{1/2}$) and the fingers spacings ($a\rm_{1/2}$).
THz radiation at 2.54~THz is applied at normal incidence.
(b) Drain-to-source current as a function of the gate voltage $U_{\rm G1}$
measured at $U_{\rm G2} =0$~V. Inset shows  the photograph of the structure. 
Here G$\rm_{1}$/G$\rm_{2}$, S and D denote first/second gate,  
source and drain, respectively.  Part of G$\rm_{1}$/G$\rm_{2}$ 
structure is highlighted by yellow lines for visualization. 
} \label{fig01}
\end{figure}

The device structure is based on an InAlAs/InGaAs/InAlAs/InP 
high-electron mobility transistor (HEMT) and incorporates doubly 
inter-digitated grating gates (DGG)  G$_1$ and G$_2$.
A sketch and a photograph of the gates are shown in 
Fig.~\ref{fig01}(a) and inset in Fig.~\ref{fig01}(b).
The 2D electron channel is formed in a quantum well (QW) at the heterointerface 
between a 16\,nm-thick undoped InGaAs composite channel layer and a 23\,nm-thick, 
Si-doped InGaAs carrier-supplying layer. 
The electron density of the 2DEG is about $3 \times 10^{12}$\,cm$^2$,  
electron effective mass normalized on free-electron mass $m_0$
and room temperature mobility are
$m/m_0\,=\,0.04$  and $\mu_0$\,=\,11000\,cm$^2$/(Vs), respectively.
The DGG gate is formed with 65\,nm-thick Ti/Au/Ti by a standard lift-off process. 
The footprint of the narrower gate fingers G$_1$ was defined by an 
E-beam lithography, whereas that of the wider gate fingers G$_2$ was 
defined by a photolithography.
In all studied structures, the metal fingers of the grating gates 
G$_{1}$ and G$_{2}$ have the same length, being 
$d_{\rm G1}$\,=\,200\,nm and $d_{\rm G2}$\,=\,800\,nm. 
The  spacing between narrow and wide DGG fingers is asymmetric with 
$a_{\rm G1}$\,=\,200\,nm  and $a_{\rm G2}$\,=\,400\,nm, see Fig.~\ref{fig01}. 
The size of the active area, covered with the grating is about 
20\,$\mu$m$\times$20\,$\mu$m.
Ohmic contacts, forming source and drain of HEMTs, 
were fabricated by highly doped 15\,nm thick InAlAs and InGaAs layers.
The axis along the gate's 
fingers  is denoted as $x$ and that along source and drain  as $y$. The characteristic source/drain current - gate voltage
dependence obtained by 
transport measurement is shown for sample\,\#A in Fig.~\ref{fig01}(b).

\begin{figure}
\includegraphics[width=0.85\linewidth]{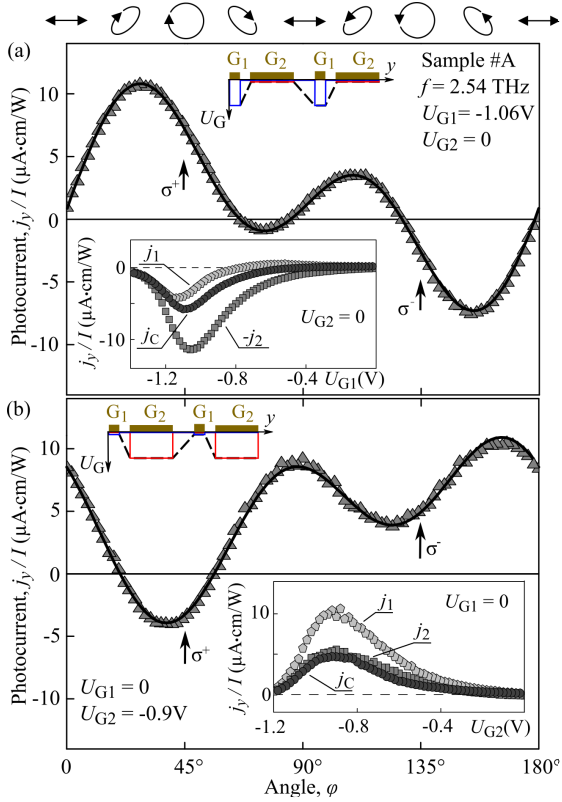}
\caption{THz radiation induced normalized photocurrent $j_y /I$ as a function of 
the angle $\varphi$ defining the radiation helicity.
The current is measured for different voltages applied to the first and second gates. 
(a) shows the data for {$U_{\rm G1}\,=\,-1.06$\,V} at gate 1
and zero gate voltage at gate 2. 
(b) shows the photocurrent measured for zero gate voltage at gate 1
and $U_{\rm G2}\,=\,-0.9$\,V. Full lines show fits to the total current 
calculated after Eq.~(\ref{FITphi}).
The ellipses on top illustrate the
polarization states for various $\varphi$.
Insets show amplitudes of  photocurrent contributions   
$j_{\rm C}/I$, driven by the light helicity,  and $j_1/I$ ($j_2/I$), induced by linear polarization, as a function of the gate voltages $U_{\rm G1}$ or $U_{\rm G2}$. 
Second set of the insets schematically show corresponding gate potentials.
Dashed lines are guide for the eye indicating the potential asymmetry 
in $y$-direction. Note that presence of the metal gates results in a 
nonzero potential even for $U_{\rm \rm G}\, =\, 0$. 
} \label{fig02} 
\end{figure}

All experiments are performed at room temperature.
The HEMT structures  were illuminated with polarized THz and microwave (MW) radiation at normal incidence.
For optical excitation we used low power $cw$ optically pumped CH$_3$OH 
THz laser~\cite{3aa,karch2010} and Gunn diodes
%
providing monochromatic radiation with frequencies 
$f$\,=\,2.54\,THz 
and  95.5\,GHz, 
respectively.
%
The radiation peak power $P$, being of the order of several milliwats at the sample's 
position, has been controlled by  pyroelectric detectors and focused 
onto samples by parabolic mirrors  (THz laser) or horn antenna (Gunn diode). 
The spatial beam distribution of THz radiation had an almost Gaussian profile, 
checked with a pyroelectric camera~\cite{edge,Glazov2014}. 
THz laser radiation peak intensity, $I$,  
for laser spot being of about 1.2~mm diameter on the sample, 
was $I \approx 8$~W/cm$^2$. 
The profile of the microwave radiation and, in particular, the efficiency of the 
radiation coupling to the sample couldn't be determined with satisfactory accuracy. 
Thus, all microwave data are given in arbitrary units.
The polarization state of THz radiation has been varied applying crystal quartz 
$\lambda$/4- or $\lambda$/2-plates~\cite{book}. To obtain circular and elliptically polarized light
the quarter-wave plate was rotated by the angle, $\varphi$, 
between the initial polarization plane and the optical axis of the plate. 
The radiation polarization states for several angles $\varphi$ are 
illustrated on top of Fig.~\ref{fig02}.
Orientation of the linearly polarized radiation is defined 
by the azimuth angle $\alpha$, 
with $\alpha\,=\,\varphi\,=\,0$  chosen in such a way that the electric 
field of incident linearly polarized light 
is directed along  $x$-direction.
Different orientation of linearly polarized MW radiation were obtained 
by rotation of a metal wire grid polarizer. 
The photocurrent excited between source and drain 
is measured across a 50 $\Omega$ load resistor 
applying the standard lock-in technique. 

\begin{figure}
\includegraphics[width=0.85\linewidth]{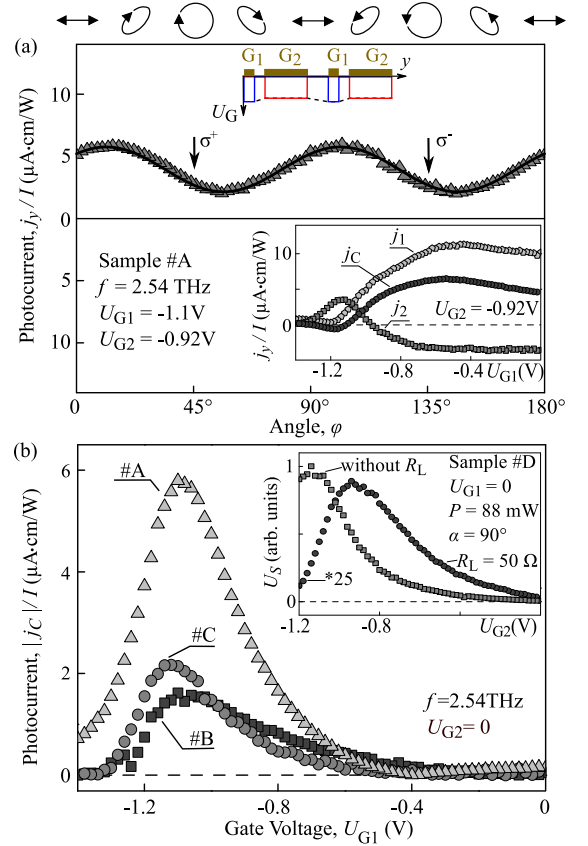}
\caption{(a) THz radiation induced normalized photocurrent $j_y /I$  as a function of 
the angle $\varphi$ defining the radiation helicity.
The current is measured for comparable voltages applied to
the first ($U_{\rm G2}\,=\,-1.1$\,V)  and the second ($U_{\rm G2}\,=\,-0.92$\,V) gates. 
Full line shows fit to the total current 
calculated after Eq.~(\ref{FITphi}).
The ellipses on top illustrate the
polarization states for various $\varphi$.
Right inset shows amplitudes of photocurrent contributions   
$j_{\rm C}/I$, driven by the light helicity,  and $j_1/I$ ($j_2/I$), induced by 
linear polarization, as a function of the gate voltage $U_{\rm G1}$ or $U_{\rm G2}$. 
Upper inset schematically shows corresponding gate potentials.
Dashed lines are guide for the eye indicating the potential asymmetry in $y$-direction.
(b) shows amplitudes of the photocurrent contributions   
$j_{\rm C}/I$, driven by the light helicity, as a function of the gate voltage $U_{\rm G1}$ ($U_{\rm G2}$\,=\,0)
measured for three different structures \#A, \#B, and \#C.
The inset shows photovoltage measured in sample \#D
across 50~$\Omega$ load resistance ($R_L   \ll R_s$) and 
directly from the sample over the lock-in amplifiers 
input resistance being much larger than the sample resistance $R_s$.
Note that the former signal is multiplied by factor 25.
} \label{fig03} 
\end{figure}

\begin{figure}
\includegraphics[width=0.85\linewidth]{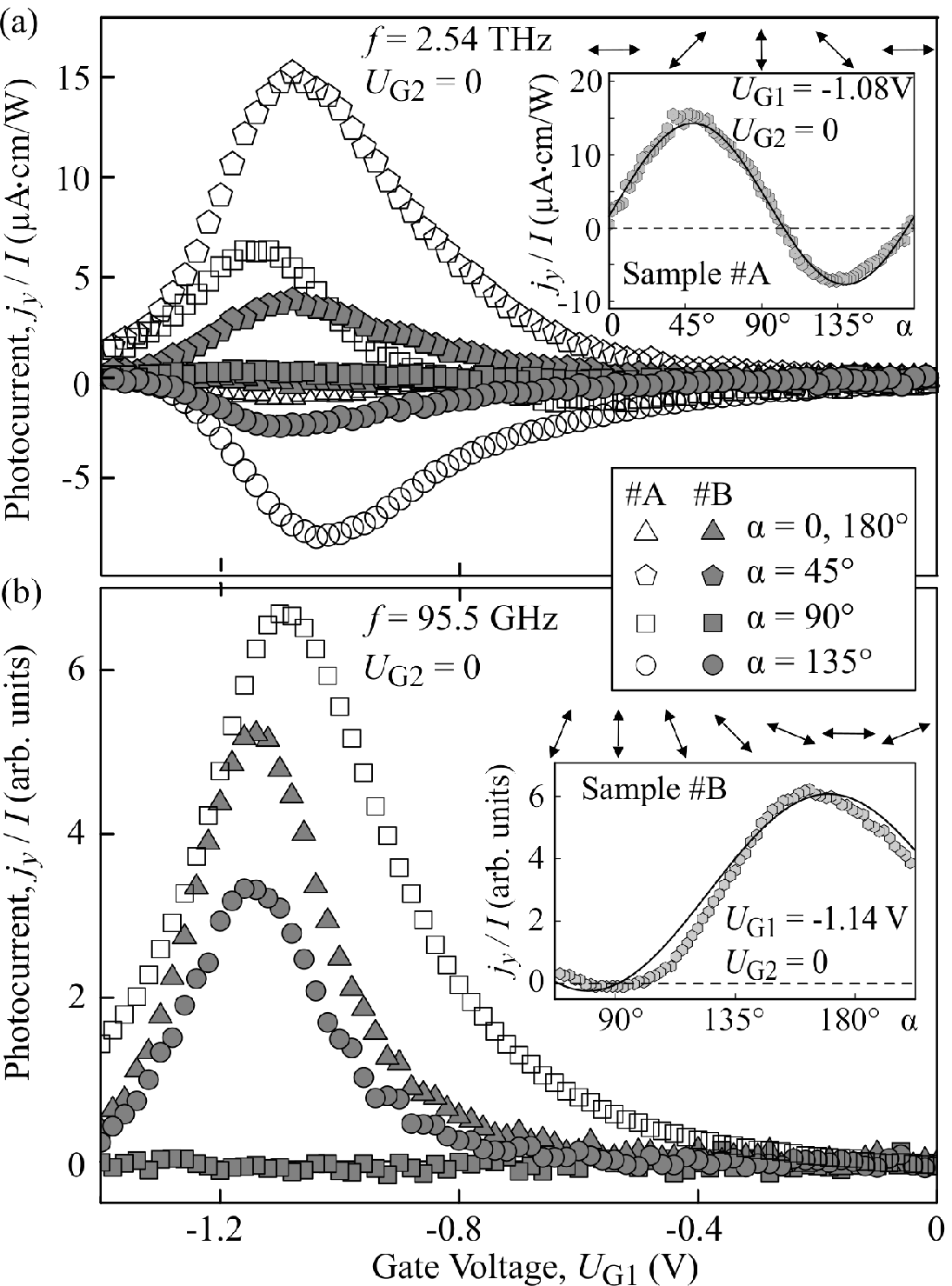}
\caption{(a) THz radiation induced normalized photocurrent $j_y/I$ 
excited by linearly polarized THz radiation in samples \#A 
and \#B as a function of the gate voltage $U_{\rm G1}$. 
The current is shown for $U_{\rm G2}\,=$\,0 and several in-plane orientations 
of the radiation electric field in respect to source-drain line defined by azimuth angles
$\alpha$. Inset shows dependence of  $j_y$ on the angle $\alpha$ obtained for $U_{\rm G1}\,=\,-1.08$~V
and $U_{\rm G2}\,=\,0$. Full line shows fit to the  total current calculated after Eq.~(\ref{SCalpha}).
Arrows indicate electric field orientation for several angles $\alpha$.
%
(b) Photocurrent $j_y/I$ excited by linearly polarized microwave radiation 
($f =95.5$\,GHz) in samples \#A and \#B as a function of the gate voltage $U_{\rm G1}$ 
($U_{\rm G2}= 0$). Inset shows dependence of  
$j_y/I$ on the azimuth angle $\alpha$ obtained in sample \#B 
for $U_{\rm G1}\,=\,-1.14$~V and $U_{\rm G2}\,=\,0$. 
Full line shows fit after $j_y \propto \cos^2(\alpha + \theta)$ 
with the phase angle $\theta$.
} 
 \label{fig04} 
\end{figure}

\section{III. Photocurrent experiment}

Illuminating the structure with elliptically (circular) 
polarized radiation of terahertz laser operating at 
frequency $f$\,=\,2.54\,THz we observed a $dc$ current 
 strongly depending on the radiation polarization.
Figure~\ref{fig02}(a) shows the photocurrent as a function of the
phase angle $\varphi$ defining the radiation 
polarization state. The data are
obtained for zero gate voltage at the gate\,2, $U_{\rm G2}$\,=\,0
and $U_{\rm G1}$\,=\,-1.06\,V.
The principal observation is that for right- ($\sigma^+$) 
and left-handed ($\sigma^-$) polarizations, i.e., for $\varphi=45^\circ$ and $135^\circ$, 
the signs of the photocurrent $j_y$ are opposite. 
The overall dependence $j_y(\varphi)$ is well described by

\begin{equation} \label{FITphi}
j_y(\varphi) = j_0 s_0 + j_1 s_1 (\varphi) + 
j_2 s_2(\varphi) + j_{\rm C} s_3(\varphi)\:,
\end{equation}

and corresponds to the superposition of the Stokes parameters 
with different weights given by the coefficients $j_0$, $j_1$, 
$j_2$, and $j_{\rm C}$, which in the experimental geometry applying rotation 
of quarter-wave plate the Stokes parameters change after 
\begin{eqnarray}
\label{S12}
&&s_0 \equiv |E_x|^2+|E_y|^2, \\
&&s_1 \equiv |E_x|^2-|E_y|^2 = \frac{\cos{4\varphi} + 1}{2}, \\
&&s_2 \equiv E_xE_y^*+E_x^*E_y = \frac{\sin{4\varphi}}{2},\\
&&s_3 \equiv i(E_xE_y^*-E_x^*E_y) = -P_{\rm circ} = -\sin{2\varphi}\,\,\, ,
%
%
\end{eqnarray}
Here $s_0$ determines the radiation intensity, $s_1$ and $s_2$ define 
the linear polarization of radiation in the $(xy)$ and rotated by $45^{\circ}$ 
coordinate frames, 
and $s_3$ describes the degree of circular polarization or helicity of radiation.
Consequently individual photocurrent contributions in Eq.~(\ref{FITphi}) are induced 
by unpolarized, linearly or circularly polarized light components.
While the polarization dependence given by Eq.~(\ref{FITphi}) 
has been detected for arbitrary relations between voltages applied to 
the first and second gates, the magnitude and even the sign of 
the individual contributions can be controlled by the gate voltages.
The inset in Fig.~\ref{fig02}(a) shows a gate dependence of 
the polarization dependent 
contributions to the total photocurrent~\cite{footnote}. 
The dependence on the gate voltage $U_{\rm G1}$ is obtained for zero biased 
second gate. Photocurrent measured in the close circuit 
configuration with $R_L \ll R_s$ shows a maximum amplitude for 
$U_{\rm G1} = -1.1$\,V. 
%
For open circuit configuration the measured photovoltage
increases at larger negative bias voltages and achieves maximum 
at the threshold voltage, $U_{\rm th} = -1.3$\,V. Corresponding data will be 
presented and discussed below.
While the non-monotonic behavior of the signal for gate voltage variation 
is well known for FET detectors~\cite{knaprev2013,knaptredicuccirev2013,30w}
the signal sign inversion upon a change of the radiation polarization, 
see~Fig.~\ref{fig02}(a),
is generally not expected for standard Dyakonov-Shur FET detectors
indicating crucial role of the lateral superlattice in the photocurrent generation. 
To demonstrate that the observed effect indeed stems from the lateral asymmetry 
of the periodic potential 
we interchanged the voltages applied to the gates. 
Figure~\ref{fig02}(b) shows the results obtained for zero gate voltage at the first gate and 
$U_{\rm G2}= -0.9$\,V at the second one. The figure reveals that changing the sign of the lateral
potential asymmetry, see insets of Fig.~\ref{fig02}(a) and (b), 
results in the sign inversion of all contributions besides the polarization independent offset. 
The situation holds for almost all values of $U_{\rm G2}$, see
the insets in Fig.~\ref{fig02}(a) and (b).
Significantly, the proper choice of the relation between
amplitudes of the individual gate potentials 
allows one to suppress completely one or the other 
photocurrent contribution. Figure~\ref{fig03}(a) 
demonstrates that for close values of gate voltages 
the circular photocurrent vanishes (corresponding potential profile 
for $U_{\rm G1}$= -1.1\,V and $U_{\rm G2}$= -0.9\,V
is shown in the inset in Fig.~\ref{fig03}). 
The interplay of the contributions upon variation of $U_{\rm G1}$  
and for fixed $U_{\rm G2}$= -1.1\,V is shown in the inset 
in Fig.~\ref{fig03}(a). 
It is seen that for nonzero second gate voltage the 
circular, $j_c$, and linear, $j_2$, 
photocurrent contributions change their direction
with increasing $U_{\rm G1}$.
Moreover, the inversions take place at different $U_{\rm G1}$ voltages.
This fact can be used to switch on and off the circular photocurrent $j_C \propto P_{\rm circ}$ contribution.

To support the conclusion that $j_{1}$ and $j_{2}$ photocurrent contribution are caused by the linear polarized 
light component
we carried out additional measurements applying linearly 
polarized light. The gate dependence of the normalized photocurrent $j_{y}/I$  
measured for samples \#A and \#B for several azimuth angles $\alpha$
are shown in Fig.~\ref{fig04}(a). The inset in this figure
presents the dependence of $j_{y}/I$ on the electric field orientation. 
The polarization dependence is well described by 
the Eq.~(\ref{FITphi}) taking into account that for linearly polarized 
light the last term vanishes and the Stokes parameters are given by
\begin{eqnarray} 
\label{SCalpha}
s_1(\alpha) 
=  \cos{2 \alpha}\:, \quad
s_2(\alpha) = 
\sin{2 \alpha} \:. \nonumber 
\end{eqnarray}
Here $\alpha=2\beta$ defines the orientation of the polarization plane 
and $\beta$ is the angle between  the initial polarization plane and the
optical axis  of the half-wave plate. 
The magnitudes and signs of the coefficients $j_0$, $j_1$, and $j_2$  used for the fit coincide with that applied for fitting 
of $\varphi$-dependencies obtained at the  same gate voltages.
These results  demonstrate that 
photocurrents $j_{1}$ and $j_{2}$ 
measured in set-up applying quarter-wave plate are 
indeed controlled by the degree of linear polarization 
of elliptically polarized radiation.

The polarization sensitive photocurrent has been observed in all studied 
devices of similar design and arbitrary relation between second and first gate potentials. 
The photocurrent  can always be well 
described by Eq.~(\ref{FITphi}).
Figure~\ref{fig03}(b) summarizes the data on 
the helicity driven photocurrent $j_{\rm C}/I$ detected in 
three HEMT structures upon change of $U_{\rm \rm G1}$ and for  $U_{\rm \rm G2} = 0$.
In all samples we detected similar dependencies of
the photocurrent characterized by close maximum 
positions but different signal magnitudes. 
The data of Fig.~\ref{fig03}(b) as well as 
circles in its inset are obtained in the close circuit configuration applying 50~$\Omega$ load resistance. The non-monotonic behavior of the photosignal measured in this geometry is caused by the interplay of the potential asymmetry, increasing with raising second gate voltage, and raising of the sample resistance for large gate voltages. For  the open circuit geometry (signal is fed to the high input impedance of lock-in amplifier) the maximum of the signal is detected for gate voltages being equal to the threshold voltage, $U_{\rm th}$, see squares in the inset in Fig.~\ref{fig03}(b).
Following Ref.~\cite{Watanabe} 
we estimate from the voltages measured in open circuit geometry 
the voltage responsivities for the signals corresponding to the photocurrents $j_2$ and $j_C$ as 
$R_v = U_s / P \times  S / S_t \approx 0.3$~V/W and 
0.15~V/W, respectively. 
Here
$P$ the total power of the source at the detector plane, $S$
radiation beam spot area, and  $S_t = 20 \times 20$~$\mu$m$^2$
transistor area. 
The voltage responsivities, being rather low as compared to that typically 
obtained for plasmonic FET detectors, indicates the necessity of further 
optimization of the structure design. 
Finally, we note that measurements applying microwave radiation show 
that for lower frequencies the polarization behavior 
changes qualitatively. Instead of the sign-alternating dependencies discussed above
the signal now varies after  $j_y \propto \cos^2 (\alpha + \theta)$, 
see inset in Fig.~\ref{fig04}(b). This observation
is in a good agreement with the Dyakonov-Shur theory~\cite{Dyakonov}
and was reported for many conventional plasmonic FET detectors, 
see e.g.~\cite{knaprev2013,knaptredicuccirev2013}.
The gate voltage dependence of the response shown in Fig.~\ref{fig04}
also reproduces well
the results previously obtained for similar structures~\cite{Watanabe,Coquillat2014}.
Even the fact that the
maximum of the signal in various structures has been 
obtained for different directions of the electric field 
vector in respect to $y$-direction (source-drain)
has already been reported for these transistors and attributed to the 
antenna coupling of MW radiation to transistor, see Ref.~\cite{Coquillat2014}. 

\section{IV. Discussion}

The observation of the circular photocurrent 
and the sign-alternating linear photocurrent $j_2$
reveals that a microscopic process actuating these photocurrents
goes beyond  the plasmonic Dyakonov-Shur model typically applied
to discuss operation of FETs THz detectors. 
Indeed, as addressed above, the latter implies an oscillating electric 
field along source-drain direction ($y$-direction) yielding
sign conserving variation upon rotation of polarization plane, 
$j_y \propto \cos^2 \alpha $~\cite{footnote2}. 
As recently shown  in Ref.~\cite{117,118}, the Dyakonov-Shur model 
in fact may result in the circular photocurrent but
only due to interference effects of two different channels and 
two interacting antennas in small size special design FETs - 
the model which can hardly be applied to the large DGG samples used in our experiments.
At the same time, the observed polarization behavior is characteristic 
for the electronic ratchet effects excited in asymmetric periodic 
structures~\cite{28,30,29,33Roj} and 
linear/circular plasmonic ratchet effects~\cite{35,36}.
The ratchet currents
arise  due to the phase shift between the periodic potential and 
the periodic light electric field resulting from near field 
diffraction in a system with broken symmetry. 
Microscopic theory developed in Ref.~\cite{29} shows that the
helicity dependent photocurrent appear because the carriers in the
laterally modulated quantum wells move in two directions and
are subjected to the action of the two-component electric field.
Symmetry analysis of the photocurrent shows that
in our DDG structures described by C$_1$ point group symmetry~\cite{footnote3}  it varies with radiation polarization after
Eq.~(\ref{FITphi}), being in agreement with experimental observation shown in Figs.~\ref{fig02},~\ref{fig03} and \ref{fig04}(a).
Moreover, as the ratchet photocurrents are proportional to the
degree of the in-plane asymmetry, they reverse the sign
upon inversion of static potential asymmetry. Exactly 
this behavior has been observed in experiment, see Fig.~\ref{fig02} (a) and (b). 
The proportionality to the degree of lateral asymmetry also 
explain the  increase of the signal with raising voltage
applied to one gate at constant voltage by the other. 
The interplay of the degree of lateral asymmetry and
periodic modulation of THz electric field
results in the complex gate-voltage dependence, in particular, for
$U_{\rm G1} \approx U_{\rm G2}$. 
As the different individual contributions to 
the total current effect might imply different microscopic mechanisms 
of the photocurrent formation,
their behavior upon change of external parameters can 
distinct from each other. 
This would result in a sign-alternating
gate-voltage behavior, in particular for the range of comparable $U_{\rm G1}$ and $U_{\rm G2}$, like 
it is observed in experiment,
see Fig.~\ref{fig02} (c). 
While all qualitative features of the observed phenomena 
can be rather good  described in terms of ratchet effects we would like to address
another possible effect, which might trigger the helicity-driven photocurrent.  
It could be the differential plasmonic drag effect in the two-dimensional structure 
with an asymmetric double-grating gate considered in Refs.~\cite{35,Popov2015}. 
As shown in Ref.~\cite{35} for a periodic AlAs/InGaAs/InAlAs/InP structure 
and linearly polarized THz radiation, photon drag effect can be comparable in strength with 
the plasmonic ratchet effect at THz frequencies. As the circular photon drag 
effect has been observed in different low dimensional materials~\cite{karch2010,Shalygin2006} 
we can expect that modification of the theory developed in~\cite{35} can 
also yield helicity driven plasmonic drag current 
compatible with the ratchet one.

Finally, we note that the ratchet effects (either electronic or plasmonic) can be 
greatly increased due to the resonant enhancement of the 
near-field in two-dimensional electron system at the 
plasmon resonance excitation as it was shown for the plasmonic ratchet 
in Refs.~\cite{31Roj,36}. The resonant plasmon condition 
$\omega  \tau >1$, see Ref.~\cite{Dyakonov} can be well 
satisfied in our structure ($\omega  \tau =4$  at 2.54~THz). 
As shown in Ref.~\cite{31Roj}, the fundamental plasmon resonance 
is excited in a similar structure at frequency around 2~THz. Therefore, 
the plasmon resonance excitation can contribute to the observed ratchet 
effects independently of particular microscopic mechanisms of the ratchet 
photocurrent formation. The measurements in a broader THz frequency range 
could elucidate the role of the plasmonic resonance excitation in the 
ratchet photocurrent enhancement.

\section{V. Summary}

To summarize, our measurements demonstrate that dual-grating-gate 
InAlAs/InGaAs/InAlAs/InP excited by terahertz radiation can yield a 
helicity sensitive photocurrent response at THz frequencies. 
We show, that HEMTs with asymmetric lateral superlattice 
of gate fingers with unequal widths and spacing can be applied  for
generation of a photocurrent defined by linearly and circularly radiation 
polarization components. Moreover, one can obtain photoresponse 
being proportional to one of the Stokes parameters simply by variation 
of voltages applied to the individual gates. The photocurrent formations
can be well described in terms of ratchet effects excited by 
terahertz radiation. By that the lateral grating induces a periodical 
lateral potential acting on the 2D electron gas in QW. This grating also modulates 
the incident radiation in the near field and hence in the plane of the 2DES, 
resulting in  circular,  linear and polarization-independent ratchet effects.
While the responsivity of the polarization dependent response is lower than that 
reported for FET transistors it can be substantially improved by 
optimization of the structure design leading 
the resonant enhancement of  the ratchet effects the plasmon resonance excitation.

\acknowledgments  We thank V. Popov for helpful
discussions.   The financial support from the DFG (SFB 689)
is gratefully acknowledged.


\begin{thebibliography}{99}

\bibitem{knaprev2013}  W. Knap and M. Dyakonov,  \textit{Plasma wave THz detectors and emitters }
in Handbook of Terahertz Technology edited by D. Saeedkia 
(Woodhead Publishing, Waterloo, Canada, 2013), pp. 121-155.

\bibitem{knaptredicuccirev2013} W. Knap, , 
S. Rumyantsev, M. S Pea Vitiello, D. Coquillat, S. Blin, N. Dyakonova, M. Shur, F. Teppe, A. Tredicucci, T. Nagatsuma, 
Nanotechnology \textbf{24}, 
214002 (2013).

\bibitem{Boppel2012} S. Boppel,  A. Lisauskas, A. Max, V. Krozer, and H. G. Roskos, 
Opt. Lett. \textbf{37}, 536 (2012).

\bibitem{Muravev2012} V. M. Muravev and I. V. Kukushkin, Appl. Phys. Lett. \textbf{100}, 082102 (2012).

\bibitem{Tonouchi2007} M. Tonouchi, 
Nature Photon.
\textbf{1}, 97 (2007).

\bibitem{9} F. Schuster, D. Coquillat, H. Videlier, M. Sakowicz, F. Teppe, L. Dussopt, B. Giffard, T. Skotnicki, and W. Knap, 
Opt. Express \textbf{19}, 7827
(2011).

\bibitem{14Roj} G. C. Dyer, S. Preu, G. R. Aizin, J. Mikalopas, A. D. Grine, J. L. Reno, J. M. Hensley, N. Q. Vinh, A. C. Gossard, M. S. Sherwin, S. J. Allen, and E. A. Shaner,
Appl. Phys. Lett. \textbf{100}, 083506 (2012).

\bibitem{11} S. Preu, M. Mittendorff, S. Winnerl, H. Lu, A. C. Gossard, and H. B. Weber, 
Opt. Express \textbf{21}, 17941 (2013).

\bibitem{Watanabe} T. Watanabe, S. A. Boubanga-Tombet, Y. Tanimoto, D. Fateev, V. Popov, D. Coquillat, W. Knap, Y. M. Meziani, Y. Wang, H. Minamide, H. Ito, and T. Otsuji, 
IEEE Sensors Journal \textbf{13}, 89 (2013).

\bibitem{But2014}  D. B. But, C. Drexler, M. V. Sakhno, N. Dyakonova , O. Drachenko, F. F. Sizov, A. Gutin, S. D. Ganichev, W. Knap, 
J. Appl. Phys. \textbf{115}, 164514 (2014).

\bibitem{Knap2014} W. Knap, D. B. But, N. Dyakonova, D. Coquillat, A. Gutin, O. Klimenko, S. Blin, F. Teppe, M.S. Shur, T. Nagatsuma, S.D. Ganichev, and T. Otsuji, 
\textit{Recent Results on Broadband Nanotransistor Based THz Detectors} in NATO Science for Peace and Security Series B, Physics and Biophysics: THz and Security Applications, edited by C. Corsi, F. Sizov, (Springer, Dordrecht, Netherlands, 2014) pp.189 - 210.

\bibitem{Dyakonov}  M. Dyakonov and M. S. Shur, 
IEEE-Trans-ED \textbf{43}(3), 380 (1996).

\bibitem{98} W. Knap, Y. Deng, S. Rumyantsev, J.-Q. Lu, M. S. Shur, C. A. Saylor and L. C. Brunel,
Appl. Phys. Lett. \textbf{80}, 3433 (2002). 

\bibitem{99} W. Knap, V. Kachorovskii, Y. Deng, S. Rumyantsev, J.-Q. Lu, R. Gaska, M. S. Shur, G. Simin, X. Hu, M. Asif Khan, C. A. Saylor, and L. C. Brunel, 
J. Appl. Phys. \textbf{91}, 9346 (2002). 

\bibitem{19w} T. Otsuji, M. Hanabe, T. Nishimura, and E. Sano, 
Opt.
Exp. \textbf{14}, 4815 (2006).

\bibitem{116} S. Sassine, Yu. Krupko,  J.-C. Portal, Z. D. Kvon, R. Murali, K. P. Martin, G. Hill, and A. D. Wieck, 
Phys. Rev. B \textbf{78}, 045431 (2008). 

\bibitem{20w}D. Coquillat , S. Nadar, F. Teppe, N. Dyakonova, S. Boubanga-Tombet, W. Knap, T. Nishimura, T. Otsuji, Y. M. Meziani, G. M. Tsymbalov, and V. V. Popov, 
Opt. Exp. \textbf{18},  6024 (2010).

\bibitem{16w} V. V. Popov, 
J. Infr. Millim. THz Waves \textbf{32}, 1178 (2011).

\bibitem{18w} G. C. Dyer, G. R. Aizin, J. L. Reno, E. A. Shaner, and S. J. Allen, 
IEEE J. Sel.
Topics Quantum Electron. \textbf{17}, 85 (2011).

\bibitem{31} V. V. Popov, D. V. Fateev, T. Otsuji, Y. M. Meziani, D. Coquillat, and W. Knap, 
Appl. Phys. Lett. \textbf{99}, 243504 (2011).  

\bibitem{115} E. S. Kannan, I. Bisotto, J.-C. Portal, T. J. Beck, and L. Jalabert, 
Appl. Phys. Lett. \textbf{101}, 143504 (2012).

\bibitem{35} V. V. Popov, 
Appl. Phys. Lett. \textbf{102}, 253504 (2013).  

\bibitem{31Roj}  V.V. Popov, D.V. Fateev, T. Otsuji, Y.M. Meziani, D. Coquillat, and W. Knap, 
Appl. Phys. Lett. \textbf{99}, 243504 (2011).

\bibitem{28} P. Olbrich, E.L. Ivchenko, T. Feil, R. Ravash, S.D. Danilov, J.Allerdings, D. Weiss, and S. D. Ganichev, 
Phys. Rev. Lett \textbf{103}, 090603 (2009). 

\bibitem{30} E. L. Ivchenko and S. D. Ganichev, 
JETP Lett. \textbf{93}, 752 (2011).

\bibitem{29} P. Olbrich, J. Karch, E. L. Ivchenko, J. Kamann, B. Maerz, M. Fehrenbacher, D. Weiss, and S. D. Ganichev, 
Phys. Rev. B \textbf{83}, 165320 (2011).

\bibitem{33Roj} A. V. Nalitov, L. E. Golub, E. L. Ivchenko, Phys. Rev.
B \textbf{86}, 115301 (2012).

\bibitem{36} I. V. Rozhansky, V. Yu. Kachorovskii, and M. S. Shur, 
Phys. Rev. Lett. \textbf{114}, 246601 (2015).

\bibitem{Saleh} B. E. A. Saleh, M. C. Teich, \emph{Fundamentals of Photonics} (John Wiley \& Sons, New York,  2003).

\bibitem{122} S. N. Danilov, B. Wittmann, P. Olbrich, W. Eder, W. Prettl, L. E. Golub, E. V. Beregulin, Z. D. Kvon, N. N. Mikhailov, S. A. Dvoretsky, V. A. Shalygin, N. Q. Vinh, A. F. G. van der Meer, B. Murdin, and S. D. Ganichev, 
J. Appl. Physics \textbf{105}, 013106 (2009). 

\bibitem{123} S. D. Ganichev, J. Kiermaier, W. Weber, S. N. Danilov, D. Schuh, Ch. Gerl, W. Wegscheider, D. Bougeard, G. Abstreiter, and W. Prettl, 
Appl. Phys. Lett. \textbf{91}, 091101 (2007). 

\bibitem{124} S. D. Ganichev, W. Weber, J. Kiermaier, S. N. Danilov, D. Schuh, W. Wegscheider, Ch. Gerl, D. Bougeard, G. Abstreiter and W. Prettl, 
J. Appl. Physics \textbf{103}, 114504 (2008). 

\bibitem{3aa} S.D.~Ganichev, S.A. Tarasenko, V.V. Bel'kov, P. Olbrich,  W. Eder,D.R.~Yakovlev, V. Kolkovsky, W. Zaleszczyk, G.~Karczewski, T. Wojtowicz, and D. Weiss, 
Phys. Rev. Lett. \textbf{102}, 156602 (2009). 

\bibitem{karch2010}
J.~Karch, P.~Olbrich, M.~Schmalzbauer, C.~Zoth, C.~Brinsteiner,   M.~Fehrenbacher, U.~Wurstbauer, M.~M. Glazov, S.~A. Tarasenko, E.~L.   Ivchenko, D.~Weiss, J.~Eroms, R.~Yakimova, S.~Lara-Avila, S.~Kubatkin, S.~D.   Ganichev,
Phys. Rev. Lett. \textbf{105}, 227402 (2010).

\bibitem{edge}
J.~Karch, C.~Drexler, P.~Olbrich, M.~Fehrenbacher, M.~Hirmer, M.~M. Glazov,   S.~A. Tarasenko, E.~L. Ivchenko, B.~Birkner, J.~Eroms, D.~Weiss, R.~Yakimova,   S.~Lara-Avila, S.~Kubatkin, M.~Ostler, T.~Seyller, S.~D. Ganichev,
  Phys. Rev. Lett. \textbf{107}, 276601 (2011). 

\bibitem{Glazov2014}  M.M.\,Glazov and S.D.\,Ganichev, 
Physics Reports \textbf{535},  101 (2014). 

\bibitem{book}S. D. Ganichev and W. Prettl, \textit{Intense Terahertz Excitation of Semiconductors}
(Oxford University Press, Oxford, 2006).

\bibitem{footnote}  While being detected in all reported measurements a polarization independent offset given by the coefficient $j_0$ 
will not be discussed in details. Instead, hereafter we focus on helicity  sensitive photocurrent, $j_{\rm C}$,
and currents driven by linearly polarized light, 
$j_{1}$ and $j_{2}$. 

\bibitem{30w} M. Sakowicz, M. B. Lifshits, O. A. Klimenko, F. Schuster, D. Coquillat, F. Teppe, and W. Knap,
J. Appl. Phys. \textbf{110},  054512 (2011).

\bibitem{Coquillat2014}
D. Coquillat, V. Nodjiadjim, A. Konczykowska, M. Riet, N. Dyakonova, C. Consejo, F. Teppe, J. Godin, W. Knap,
Didgest of Int. Conf. on Infrared, Millimeter, and Terahertz Waves, Tucson, USA, (2014). 

\bibitem{footnote2} Note that signal variation with polarization is, apart the offset, identical with that of $s_1$, therefore this polarization 
dependence can also be used to describe the $j_1$-related photocurrent behavior.

\bibitem{117} C. Drexler, N. Dyakonova, P. Olbrich, J. Karch, M. Schafberger, K. Karpierz, Yu. Mityagin, M. B. Lifshits, F. Teppe, O. Klimenko, Y. M. Meziani, W. Knap, and S. D. Ganichev, 
J. Appl. Physics \textbf{111}, 124504 (2012). 

\bibitem{118} K. S. Romanov and M. I. Dyakonov, 
Appl. Phys. Lett. \textbf{102}, 153502 (2013).

\bibitem{footnote3} All previous works aimed to the radiation
induced ratchet effects discuss the  case of unconnected  parallel 
metal stripes: a system  belonging to C$_s$ point group symmetry 
consisting of the identity element
and the reflection in the plane perpendicular to the stripes~\cite{28,30,29,33Roj,35,36}.
For this symmetry circular photocurrent $j_C$ and the photocurrent $j_2$
can be generated along stripes only 
whereas polarization independent offset and
$j_0$ and photocurrent $j_1$ are allowed in 
the perpendicular to that direction (source-drain). 
Design of our DDG structures with interconnected metal stripes 
in each of gates excludes reflection plane reducing the point group symmetry 
to C$_1$. As a result the symmetry does not imply any restrictions
and the photocurrent includes all four individual contributions ($j_0$, $j_1$, $j_2$ and $j_C$) which are allowed in any in-plain  direction. More details on the symmetry analysis of photocurrents
in quantum wells of C$_1$ symmetry can be found in~\cite{Wittmann,Belkov2008}.

\bibitem{Wittmann}
B.~Wittmann, S.N.~Danilov,  V.V.~Bel'kov, S.A.~Tarasenko, E.G.~Novik, H.~Buhmann, C.~Br\"{u}ne, L.W.~Molenkamp, E.L.~Ivchenko, Z.D. Kvon, N.N. Mikhailov, S.A. Dvoretsky, N.\,Q.\,Vinh, A.\,F.\,G.~van~der~Meer, B.~Murdin, and S.D.~Ganichev
Semicond. Sci. and Technology \textbf{25}, 095005 (2010).

\bibitem{Belkov2008} V.V. Bel'kov,  and S.D.~Ganichev,
Semicond. Sci. Technol. \textbf{23}, 114003 (2008). 

\bibitem{Popov2015}  V.V. Popov, D.V. Fateev, E.L. Ivchenko, and S.D. Ganichev, Phys. Rev. B \textbf{91}, 235436 (2015).

\bibitem{Shalygin2006} V.A. Shalygin, H. Diehl, Ch. Hoffmann,  S.N. Danilov, T. Herrle, S.A. Tarasenko, D. Schuh, Ch. Gerl, W. Wegscheider, W. Prettl and S.D. Ganichev,
JETP Lett. \textbf{84}, 570  (2006).

\end{thebibliography}
\end{document}